\documentclass[10pt,letterpaper,twocolumn,aps,prb,citeautoscript,superscriptaddress]{revtex4}

\usepackage{amsmath}    % need for subequations
\usepackage{bbm}

\usepackage{ifpdf}  %% Detects pdflatex or latex
\usepackage{graphicx}
\usepackage{units}
\usepackage{xspace}
\usepackage{txfonts}
\usepackage{bm}
\usepackage[usenames,dvipsnames]{color}

\newcommand{\moire}{moir\'{e}\xspace}
\newcommand{\cface}{SiC($000\bar{1}$)\xspace}

\newcommand{\beq}{\begin{equation}}
\newcommand{\eeq}{\end{equation}}
\newcommand{\vecb}[1]{\ensuremath{\boldsymbol{#1}}}

\ifpdf
 \DeclareGraphicsExtensions{.pdf,.png,.jpg}
\else
 \DeclareGraphicsExtensions{.eps}
\fi

\begin{document}

%opening
\title{Structural Determination of Multilayer Graphene via Atomic Moir\'{e} Interferometry}
\author{David L. Miller}
 \affiliation{School of Physics, Georgia Institute of Technology, Atlanta GA, 30332}
\author{ Kevin D. Kubista}
 \affiliation{School of Physics, Georgia Institute of Technology, Atlanta GA, 30332}
\author{Gregory M. Rutter}
 \affiliation{Center for Nanoscale Science and Technology, National Institute of Standards and Technology, Gaithersburg MD, 20899}
\author{Ming Ruan}
 \affiliation{School of Physics, Georgia Institute of Technology, Atlanta GA, 30332}
\author{Walt A. de Heer}
 \affiliation{School of Physics, Georgia Institute of Technology, Atlanta GA, 30332}
\author{Phillip N. First}
 \affiliation{School of Physics, Georgia Institute of Technology, Atlanta GA, 30332}
\author{Joseph A. Stroscio}
 \affiliation{Center for Nanoscale Science and Technology, National Institute of Standards and Technology, Gaithersburg MD, 20899}

\begin{abstract}
Rotational misalignment of two stacked honeycomb lattices produces a moir\'{e} pattern that is observable in scanning tunneling microscopy as a small modulation of the apparent surface height. This is known from experiments on highly-oriented pyrolytic graphite.  Here, we observe the combined effect of three-layer moir\'{e} patterns in multilayer graphene grown on SiC ($000\bar{1}$). Small-angle rotations between the first and third layer are shown to produce a ``double-moir\'{e}'' pattern, resulting from the interference of  moir\'{e} patterns from the first three layers. These patterns are strongly affected by relative lattice strain between the layers. We model the moir\'{e} patterns as a beat-period of the mismatched reciprocal lattice vectors and show how these patterns can be used to determine the relative strain between lattices, in analogy to strain measurement by optical moir\'{e} interferometry.
\end{abstract}

\maketitle

\section{Introduction}
    Perfect graphene is a single atomic layer of carbon atoms arranged into two interpenetrating triangular sublattices (A \& B). It has a unique linear band structure stemming from the unperturbed $\pi$ orbitals, which lie above the plane of hybridized $sp^{2}$ bonds\cite{Wallace47}. When placed on a substrate, the band structure may be modified by interaction of the $\pi$ orbitals with the substrate. One way to minimize this interaction is to use graphene as its own substrate. In Bernal stacked (AB) bilayer graphene, the low-energy band structure near the Fermi energy is not linear due to a breaking of the sublattice symmetry; however, if the layers are rotated away from Bernal stacking the sublattice symmetry and linear band dispersion is preserved over the unit cell of the newly formed \moire superlattice\cite{Hass08}.

    In addition to affecting the electronic structure, the \moire pattern resulting from lattice misalignment or mismatch can itself be a useful tool for understanding structural properties.  Optical \moire interferometry has been applied in strain analysis for many years\cite{Bromley55}. Because a scanning tunnneling microscope (STM) can detect \moire patterns in rotated graphene layers\cite{Hass08b}, analogous methods can be applied to measure local strains between individual sheets of graphene at the nanometer scale. Here we show how STM-based ``atomic \moire interferometry'' can go beyond surface properties to detect lattice orientations and strains for depths of several layers. We expect this type of analysis to be beneficial to understanding electronic and transport properties in graphene multilayers.

    Over the past 20 years, several studies of \moire patterns occurring on highly-oriented pyrolytic graphite (HOPG) have been conducted\cite{Kuwabara90,Xhie93,Rong93,Rong94,Sun03,Pong05a}. These typically occur due to a rotation between two layers near the surface, or from an exfoliated flake resting on the HOPG surface \cite{Gan03}.  Graphene islands grown on transition metal surfaces (e.g., Ir($111$), Ru($0001$), or Pd($111$)\cite{Marchini07,NDiaye08,Kwon09}) also display \moire-style superlattices due to both rotational misalignment and the inherent size mismatch of the lattices.  More recently, it was found that multilayer epitaxial graphene (MEG) grown on \cface stacks in such a way that almost every pair of graphene sheets is rotated with respect to its neighbor(s). Graphitic Bernal stacking of the graphene layers appears to be rare in this material, although there are clearly preferred layer orientations\cite{Hass08,Hass08b,Sprinkle09a}. X-ray diffraction and low-energy electron diffraction experiments show broad peaks corresponding to rotations of $\pm2.2^{\circ}$ and $30^{\circ}$ to the SiC lattice \cite{Hass08,Hass08b}. The rotational disorder of the graphene samples produces \moire patterns in STM topographs, and has been under active investigation \cite{Naitoh03,Hass08,Hass08b,Varchon08,Biedermann09,Hiebel08}. The patterns appear as an additional corrugation in STM imaging; however, the apparent height variation is thought to be dominated by a modulation in the local density of states\cite{Rong93,Campanera07}. The local stacking changes continuously between regions that resemble AA, AB, BA and slip stacking \cite{Rong93}. Despite regions of locally AB stacking, the sublattice symmetry is preserved over the full \moire unit cell. In each unit cell, AB regions are complemented by BA regions that restore the inversion symmetry. The result is that the electronic properties of an $N$-layer film of MEG are nearly equivalent to $N$ independent graphene layers: the linear band dispersion near the charge neutrality (Dirac) points of the Brillouin zone is preserved\cite{Miller09,Sprinkle09a,Hass08}.

\section{Experimental Methods}
    Multilayer graphene is grown on \cface by a low-vacuum induction furnace technique\cite{Berger04,Berger06}. When heated, the SiC crystal thermally decomposes leaving behind carbon atoms to reform into graphene sheets. The sample studied has an average of $10 \pm 1$ layers as measured by ellipsometry. It was initially imaged using atomic force microscopy (AFM) and inspected with low energy electron diffraction (LEED) and Auger electron spectroscopy (AES) to ensure the sample quality. The sample was then placed into a custom built ultra-high vacuum chamber (base pressure $<\unit[10^{-9}]{Pa}$), where it was subsequently heated to $1250$ $^{\circ}C$ in order to clean the surface after exposure to air. After cleaning, it was studied at low temperature ($\unit[4.3]{K}$) in a cryogenic STM chamber \cite{Miller09}.

\section{Multilayer Moir\'{e}}

    \begin{figure*}
      \centering
      \includegraphics[width=0.9\textwidth]{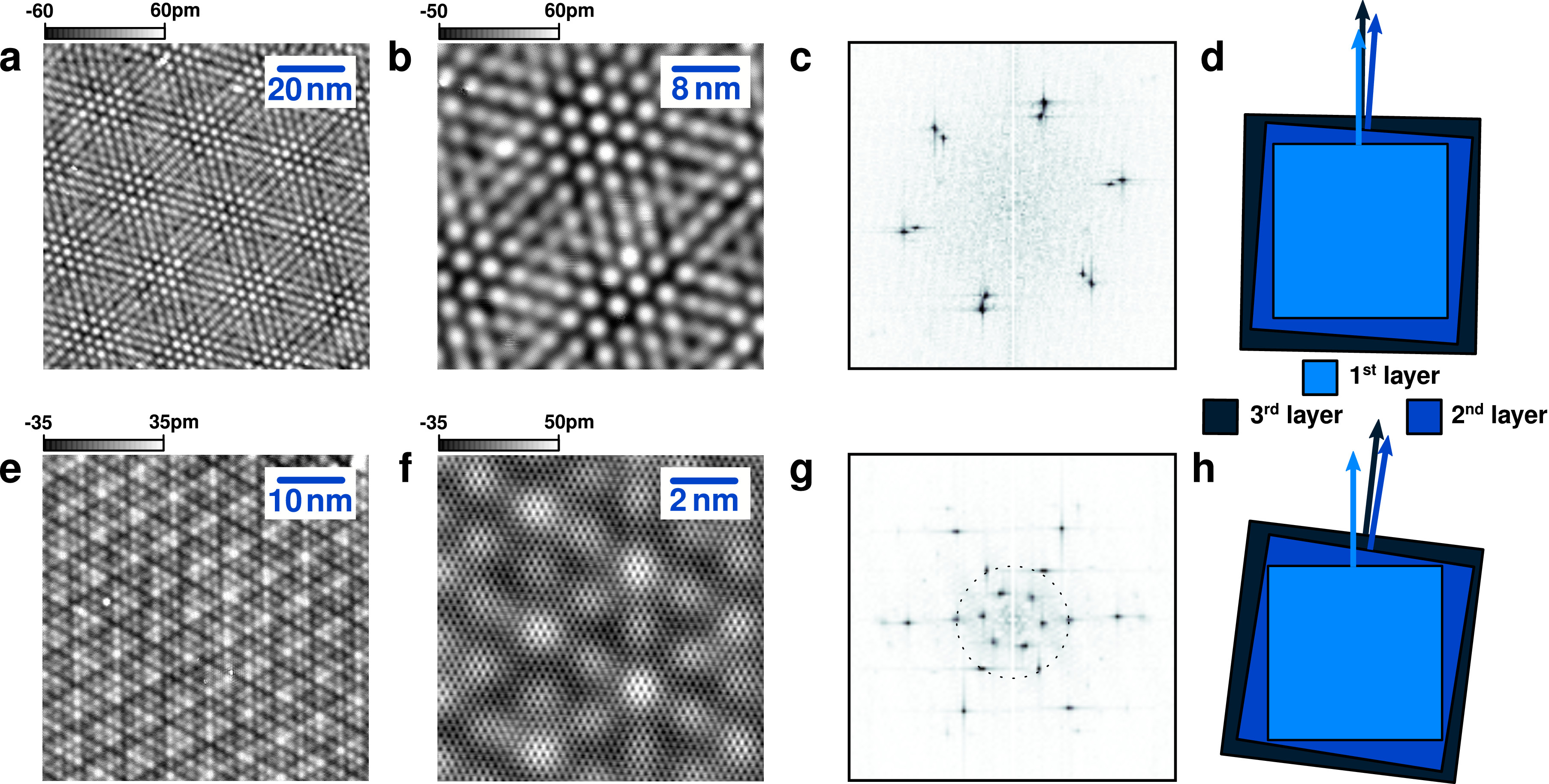}
      \caption %
      {Comparison of two \moire patterns in MEG showing multiple layer effects. (a) An image of two similar sized \moire patterns interfering, resulting in a large unit cell pattern. Layers 1 (Surface Layer) and 2, and 2 and 3 have comparable rotation angles in opposite directions (shown in d). (b) zoomed in view of (a). (c) FT of (a). (d) schematic of layer orientation based on the \moire pattern in (a). (e) A different superstructure that indicates low rotation angles between layers 2 and 3. The spots in the FT relative to the reciprocal lattice vectors were used in this case to determine which spots correspond to the rotation between layer 1 and 2. (f) zoomed in view of (e). (g) FT of (e). (h) schematic of layer orientation based on the \moire pattern in (e).}
      \label{fig:MoireComp}
    \end{figure*}

    Multilayer graphene often yields more complicated \moire patterns than HOPG, resulting from the successive rotation of layers throughout the depth of the material. Multiple \moire patterns are often seen during imaging (Figure \ref{fig:MoireComp}). Unlike HOPG, Bernal stacking does not frequently occur; thus, subsurface layers will also have \moire alignment, which can be observed. Various publications have reported an attenuation factor (AF) associated with overlayer coverage of a \moire pattern in HOPG\cite{Rong94,Pong05b,Sun03}. A fit of this data yields $AF_{n} = e^{0.81 n}$, where $n$ is the number of overlayers\cite{Pong05a}. In view of this, a second/third layer \moire pattern would be visible ($AF_{1} = 2.25$) if present but the visibility of patterns from deeper layers decays exponentially.

    The length of a superlattice cell can be evaluated as the wavelength of an interference pattern resulting from the misaligned reciprocal lattice vectors. The difference between a reciprocal lattice vector $\vecb{k}_{i}$ and a reciprocal lattice vector rotated by angle $\theta$, $\vecb{k}_{i}^{\prime} = \hat{R}(\theta)\vecb{k}_{i}$, is used to determine the length ($D$) and orientation ($\phi$) of the pattern. Since we are interested in long wavelength \moire patterns (rotation angles, $\theta \leq 5^{\circ}$), the only relevant vectors of the interference pattern will be $\Delta \vecb{k} = \vecb{k}_{i}-\vecb{k}_{j}^{\prime}$ for $i=j$ where $i$ is any of six reciprocal lattice vectors. For unstrained lattices, only one pair of vectors $\vecb{k}$ and $\vecb{k}^{\prime}$ needs to be considered because of symmetry. Thus, for two rotated lattices of undistorted length, we obtain the equation for the size of the \moire unit cell length\cite{Kuwabara90,Amidror03}
    \beq
      D = \frac{2 \pi}{|\vecb{k}-\vecb{k}^{\prime}|} = \frac{a}{2 \sin{(\theta/2})},
    \eeq
    and orientation with respect to the atomic lattice,
    \beq
      \phi=\frac{\pi}{6}-\theta/2.
    \eeq
    Figure \ref{fig:MoireComp}a shows an example of two interfering \moire patterns (the collective interference of three graphene layers), obtained by STM measurements. On first glance, one might think that the \moire pattern constitutes a significant modulation in the tip height. In fact, the graphene lattice is extremely smooth, such that any small changes in tip height become the foremost feature. Height modulations due to \moire patterns are typically only $\unit[0.05]{nm}$ peak-to-peak. From Fig.~\ref{fig:MoireComp}a and Fig.~\ref{fig:MoireComp}b, two periodicities can be seen. The large periodicity has a length of $D \approx \unit[26.5]{nm}$ while the smaller periodicity has a length of $D \approx \unit[3]{nm}$, still much larger than the atomic lattice spacing. Because more than a single \moire pattern is observed, we can conclude that this is a combined effect of at least three layers (in another area of the sample, 3 coexisting \moire patterns were found, implying the participation of at least 4 graphene layers). From the discrete fast Fourier transform (FT) in Fig.~\ref{fig:MoireComp}c, we find that two sets of spots are present in reciprocal space. These spots correspond to \moire lengths of $D_{1}=\unit[2.95 \pm 0.05]{nm}$ and $D_{2}=\unit[3.35 \pm 0.09]{nm}$\cite{Error}.

    Using equations (1) and (2) and $a = \unit[0.246]{nm}$ as the graphene lattice constant, we deduce the rotation angles needed to produce a \moire of size $D_{1}$ and $D_{2}$ as $\theta_{1} = \unit[4.78 \pm 0.07]{^{\circ}}$ and $\theta_{2} = \unit[4.21 \pm 0.11]{^{\circ}}$. Because $\Delta \phi = \phi_{1} - \phi_{2} \approx \unit[0]{^{\circ}}$, as seen in the Fourier transform, we propose the orientation of the top three layers in Fig.~\ref{fig:MoireComp}d. The first and third layers (starting from the surface and counting into the bulk) must be nearly aligned in order to obtain the observed $\Delta \phi$ in the FT. The measured value is $\Delta \phi = \unit[0.61 \pm 0.23]{^{\circ}}$, slightly higher than the expected value of $\Delta \phi = \unit[0.29]{^\circ}$ from the proposed three layer orientation (Fig.~\ref{fig:MoireComp}d) and equation (2)\cite{Error}. Again, we can determine the length of the largest interference pattern from $\Delta \vecb{k}$ using the \moire reciprocal lattice vectors from the FT,
    \beq
      D_{3} = \frac{D_{1}D_{2}}{\sqrt{D_{1}^{2}+D_{2}^{2}-2D_{1}D_{2}\cos{(\Delta \phi)}}}.
    \eeq
    Thus, we calculate a value of $D_{3}=\unit[24.7 \pm 1.6]{nm}$ for the size of the largest observed pattern\cite{Error}. The uncertainty is due to a distortion in the hexagonal shape of the pattern, and is likely caused by strain, which will be discussed later. As shown in the schematic in Fig.~\ref{fig:MoireComp}d, the top layer is nearly aligned with the third. This creates two patterns of nearly equal size and orientation. The angle between the two \moire patterns is small, producing a large interference pattern.

    In the case of Fig.~\ref{fig:MoireComp}e, we have applied a similar analysis to determine the orientation of the top three lattices. The smallest $\vecb{k}$ vectors in Fig.~\ref{fig:MoireComp}f were determined to correspond to the \moire pattern of the first and second layers by comparing the \moire pattern orientation with respect to the reciprocal lattice vectors. We propose that there is a small rotation between the second and third layer. The interference of these two \moire patterns is responsible for the hexagonal set of spots on the dotted circle in the FT (Fig.~\ref{fig:MoireComp}g). Consequently, the image in Fig.~\ref{fig:MoireComp}e looks qualitatively very different than the one in Fig.~\ref{fig:MoireComp}a.

    As further proof of concept, we image the pattern at multiple sample biases for a constant tunneling current. Because the imaging is done in a constant current mode, the tip will move closer to the surface as the sample bias ($V_{S}$) is decreased. This technique is used in imaging Bernal stacked graphene bilayers, and shows a transition in the imaging of the atomic lattice \cite{Rutter08}. At higher biases, every atom is imaged; however, as the bias is decreased the tip instead images every other atom.

    \begin{figure}
      \centering
      \includegraphics[width=0.45\textwidth]{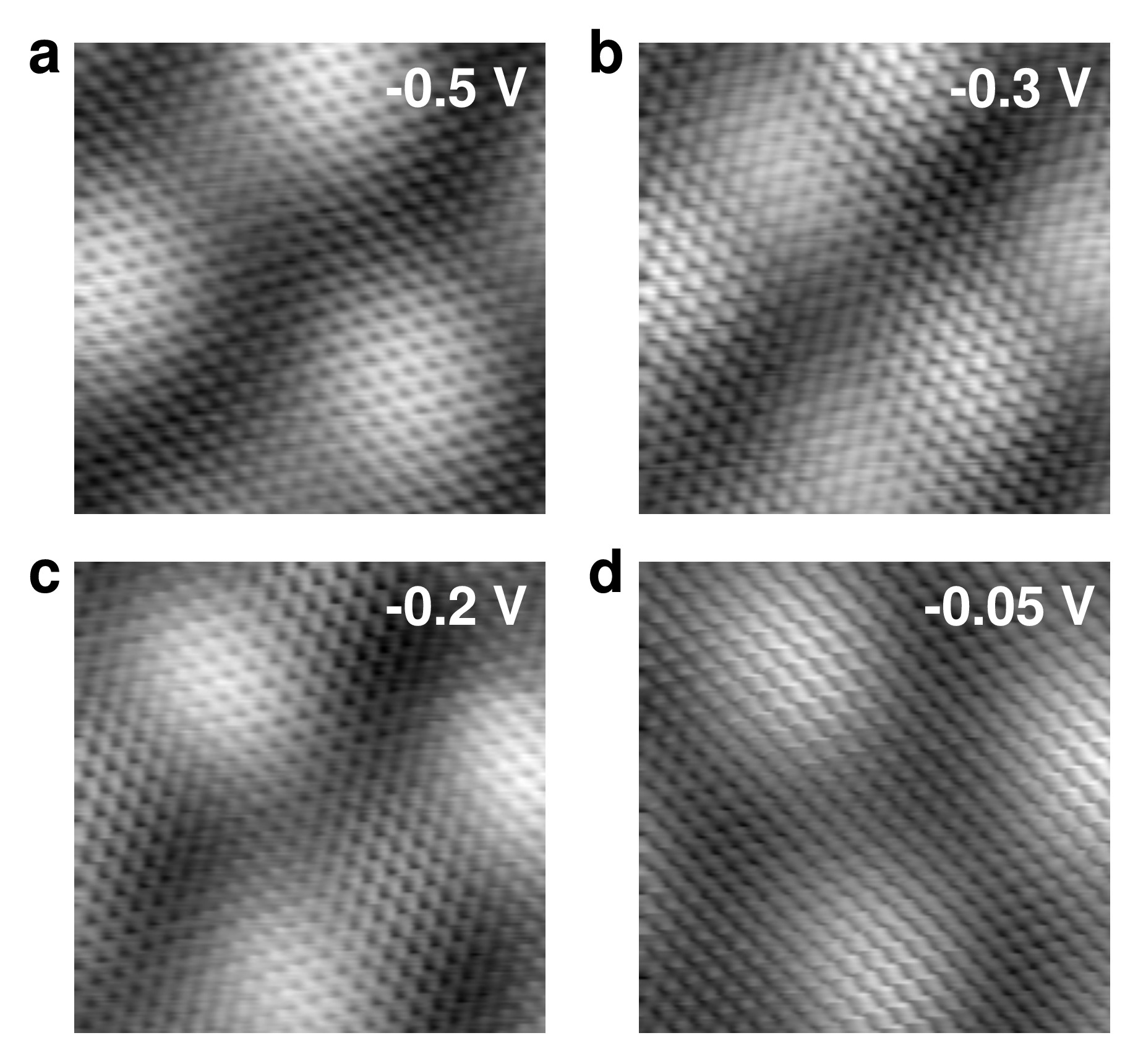}
      \caption %
      {Multibias imaging on the \moire pattern in Figure \ref{fig:MoireComp}a. Each image is $\unit[5]{nm} \times \unit[5]{nm}$, and was taken at the same spatial location and tunnel current ($\unit[100]{pA}$). In (a), only the \moire pattern of the top two layers is imaged. (b), (c) show the transition as the tip moves closer to the surface. In (d), the tip has pushed in far enough to only image the lower layer \moire pattern. We believe that by $V_{S}=\unit[-0.05]{V}$, the tip is actually in contact with the surface.}
      \label{fig:MoireMultibias}
    \end{figure}

    Figures \ref{fig:MoireMultibias}(a-d) show the result of imaging at multiple biases on a region within the image shown in Fig.~\ref{fig:MoireComp}a. At $V_{S}=\unit[-0.5]{V}$, a single set of \moire maxima is seen, corresponding to a \moire pattern from the rotation of the first/second layer. As the bias decreases, the tip begins to image a second \moire pattern. By $V_{S}=\unit[-0.05]{V}$, the tip images only the second \moire pattern, which apparently results from rotation of layers 2 and 3. From measured deviations in the expected exponential decay of tunneling current vs tip-height ($I$ vs $z$), we believe that by $V_{S}=\unit[-0.05]{V}$ the tip is in contact with the surface. The difference in tip height between $V_{S}=\unit[-0.5]{V}$ and $V_{S}=\unit[-0.05]{V}$ is $\Delta z = \unit[0.39]{nm}$. The sample biases at $V_{S}=\unit[-0.3]{V},\unit[-0.2]{V}$ (Fig.~\ref{fig:MoireMultibias}b,c) show the transition in imaging the top \moire to the one underneath. The tip bias, tunneling current and tip shape are known to have a large effect on the imaging of a \moire pattern\cite{Pong05a}. This behavior was calculated for a rotated layer on top of AB stacked graphite \cite{Cisternas08} and effects similar to our own observations were predicted; however, at present we do not understand how the \moire patterns of this multilayer system depend on tip-sample distance.

\section{Strain}

    \begin{figure}
      \centering
      \includegraphics[width=0.4\textwidth]{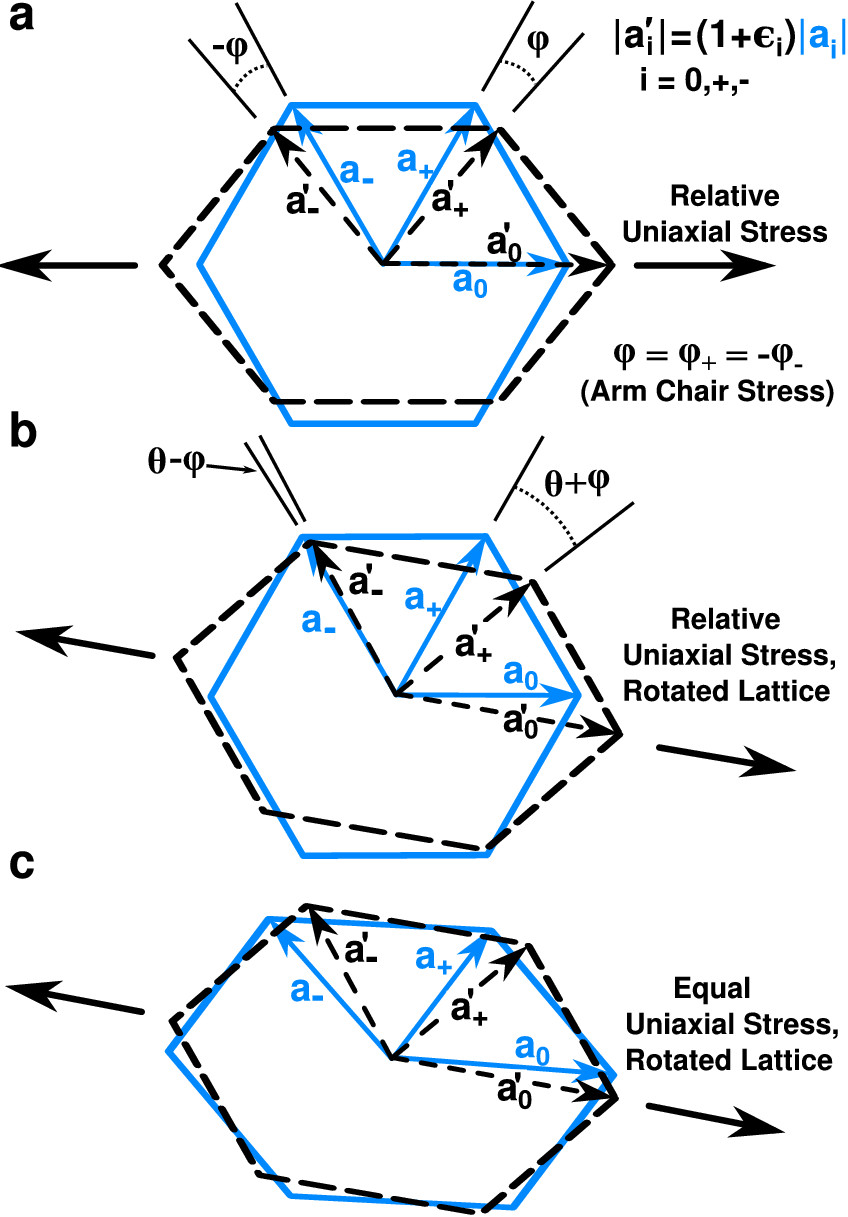}
      \caption %
      {Strain induced between two stacked hexagonal (real space) lattices in the case where one lattice (black/dashed line) is (a) stretched but not rotated and (b) stretched and rotated with respect to the undistorted lattice (blue/solid line). (c) displays equal strain for both lattices. For rotated lattices under relative strain, the angular change in the lattice vectors affect the interference pattern differently for each direction. As a result, the \moire pattern observed in STM will appear stretched or distorted. Significantly higher strain is required for similarly sized distortions if the lattices are equally strained along the same direction. At the low \moire angles ($\theta$), the lattice vectors (when equally strained) will obtain similarly sized angular distortions $\varphi_{i}$ which will nullify each other.}
      \label{fig:Strain}
    \end{figure}

    Relative strain induced between two stacked hexagonal lattices can have a large effect on observed superlattice wavelengths. This phenomenon is well known in \moire interferometry as a technique for strain measurement\cite{Bromley55}, and has been observed in large period \moire patterns on HOPG \cite{Xhie93}. The effect appears as a distortion in the real space hexagonal shape of the superlattice cell, resulting from unequal distortion of both the relative length and angular alignment of the reciprocal lattice vectors. Distorted \moire patterns are commonly observed in HOPG when lattice rotation angles ($\theta$) are small. We examine the effect of a uniform strain in the top layer relative to an unstrained layer below. For simplicity, we demonstrate the effect of a relative strain in the arm chair direction on the observed superlattice periodicity.

    Strain breaks the symmetry of the reciprocal lattice vectors, so we now consider the case where the lattice has been uniformly strained along the arm chair direction, such that $|\vecb{a}_{i}^{\prime}| = (1+\epsilon_{i})|\vecb{a}_{i}|$, where $i=0,\pm$. The vector $\vecb{k}_{0}$ along the direction of strain does not change angular orientation ($\varphi_{0} = 0$), so under rotation 
    \beq
      \vecb{k}_{0}^{\prime} = \frac{1}{(1+\epsilon_{0})}\hat{R}(\theta)\vecb{k}_{0}.
    \eeq
    This leads to a change in the wavelength of the interference pattern along the strain direction as given by the law of cosines,
    \beq
      D_{0} = \frac{2 \pi}{|\vecb{k}_{0}-\vecb{k}_{0}^{\prime}|} = \frac{a(1+\epsilon_{0})}{2 \sqrt{(1+\epsilon_{0})\sin^2{(\theta/2)}+\epsilon_{0}^{2}/4}},
    \eeq
    Lattice vectors ($\vecb{a}_{\pm}$) which are not along the direction of strain will undergo an additional rotation, $\varphi_{\pm}$. In the most general case of arbitrary changes in length ($\epsilon_{i}$) and arbitrary rotations ($\varphi_{i}$), this becomes:
    \beq
      D_{i} = \frac{2 \pi}{|\vecb{k}_{i}-\vecb{k}_{i}^{\prime}|} = \frac{a(1+\epsilon_{i})}{2 \sqrt{(1+\epsilon_{i})\sin^2{[(\theta + \varphi_{i})/2]}+\epsilon_{i}^{2}/4}}.
    \eeq
    where $i=0,\pm$, and the distortion in the angular orientation of the \moire is given by
    \beq
      \phi_{i} = \arcsin{\left(\frac{(1+\epsilon_{i}) \sin{(\theta + \varphi_{i})}}{2\sqrt{(1+\epsilon_{i})\sin^2{[(\theta + \varphi_{i})/2]}+\epsilon_{i}^{2}/4}}\right)}
    \eeq
    In order to approximate the values of $\epsilon_{i}$ for a uniform uniaxial stress, we make use of Poisson's ratio $\nu$ and the strain matrix $\vecb{\varepsilon}$ by, $\epsilon_{x}=\varepsilon,\epsilon_{y}= - \nu \varepsilon$ \cite{Kittel96}. The strain matrix is:
    \beq
      \vecb{\varepsilon} = \left( \begin{array}{cc}
      1+\varepsilon & 0 \\
      0 & 1 - \nu \varepsilon \\
      \end{array} \right),
    \eeq
    where the axes have been chosen to diagonalize the strain matrix. For strain in the arm chair direction, we do not need to rotate the lattice vectors $\vecb{a}_{i}$ with respect to $\vecb{\varepsilon}$. In this case the lattice is stressed along $\vecb{a}_{0}$. We can relate the strain matrix to the values $\epsilon_{i}$ by applying the original definintion,
    \beq
      |\vecb{a}_{i}^{\prime}| = (1+\epsilon_{i})|\vecb{a}_{i}| = |\vecb{\varepsilon}\vecb{a}_{i}|.
    \eeq 
    Thus, under a small strain approximation ($\varepsilon \ll 1$) we obtain
    \beq
	\epsilon_{0} = \varepsilon,\epsilon_{\pm} \approx \frac{1}{4}(1 - 3 \nu) \varepsilon.
    \eeq
    \beq
	\varphi_{0} = 0,\varphi_{\pm} \approx \pm \sqrt{3} \varepsilon \left| \frac{1 + \nu}{4 + \varepsilon (1 - 3 \nu)} \right|
    \eeq

    \begin{figure}
      \centering
      \includegraphics[width=0.4\textwidth]{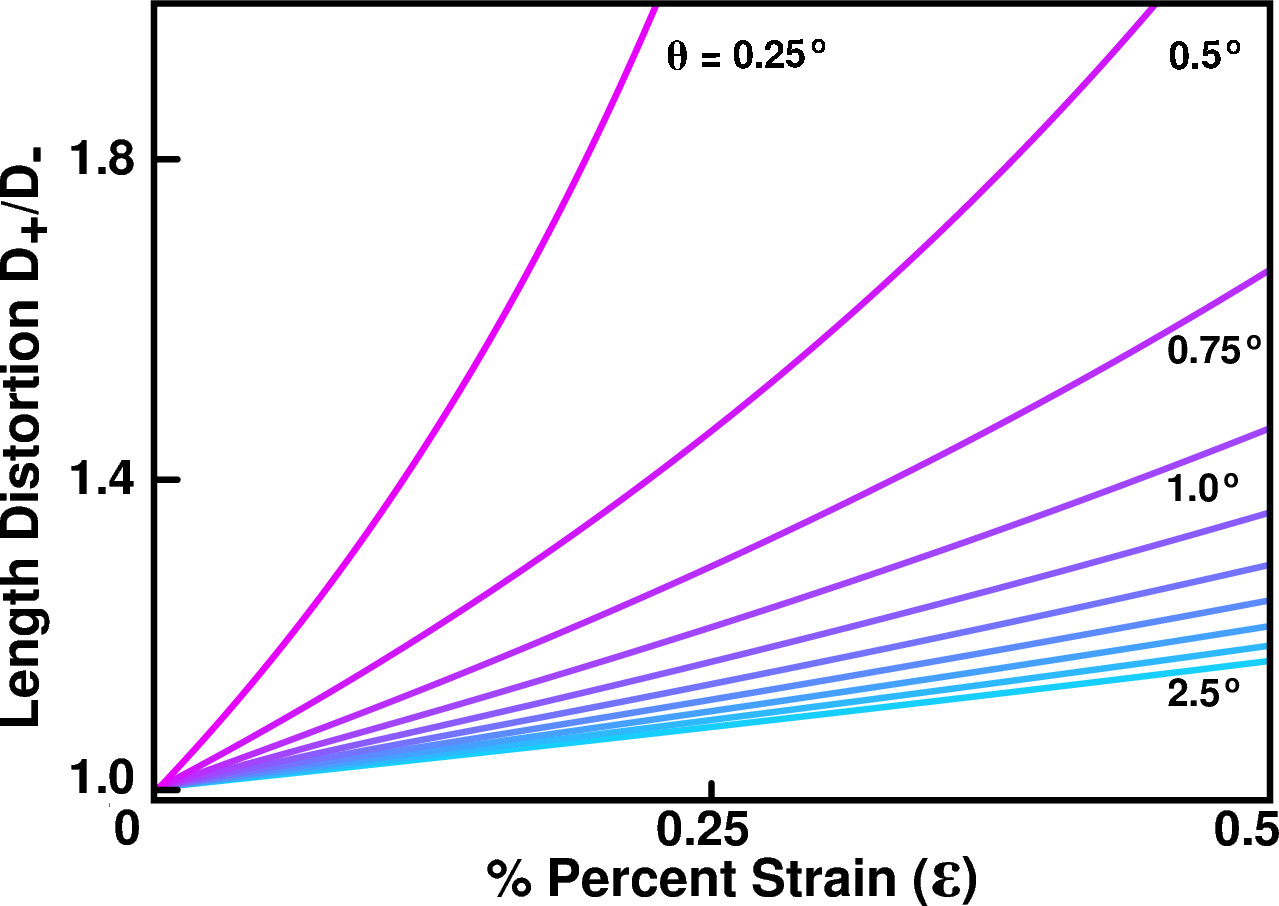}
      \caption %
      {Curves display the length distortion between two vectors of the \moire cell as a function of the \moire rotation angle $\theta$. The amount of distortion for a small strain rapidly increases with decreasing \moire rotation angles (i.e. when the rotation angle is comparable to the angular distortion, $\theta \sim \varphi$).}
      \label{fig:Distortion}
    \end{figure}

    We use the values from equations (10) and (11), along with $\nu = 0.5$ to plot the overall distortion in the wavelength of the interference pattern. If we put these values into (6), we can compare the length distortion ($D_{+}/D_{-}$) of the resulting \moire pattern. Figure \ref{fig:Distortion} clearly shows the expected behavior. As the rotation angle $\theta$ between the lattices increases, the effect of the distortion is quickly diminished; however, for a large \moire pattern the rotation angle is small, leaving the pattern easily distorted by the relative strain. The reason is that small angular differences become comparable to the rotation angle ($\varphi \approx \theta$), and lattice distortions cannot be neglected. Additionally, in this regime small changes in the lattice constant $a$ contribute greatly to the \moire rotation angle. The shape of the \moire hexagon suffers additional skewing in $\phi_{i}$ due to the angular distortion of the reciprocal lattice vectors.

    Obtaining an exact measure of the strain is difficult because it can be hard to determine the direction of the applied stress. Distortions can also be made from equally stretching both lattices along one direction as in Fig.~\ref{fig:Strain}c; however the distortions of the resulting \moire patterns are generally much smaller than for relative strains (on the order of several percent strain for a comparable length distortion). The reason is that for small \moire rotation angles, the lattice vectors will obtain similarly sized angular distortions for each lattice vector pair, $i$, thereby yielding a relative angular distortion of $\varphi_{i}-\varphi_{i}^{\prime} \approx 0$. In this case, we can use the original \moire equations to approximate the length distortion simply by replacing the lattice spacing $a$ in equation (1) with its distorted values, $a_{i} = (1+\epsilon_{i})a$.

    \begin{figure}
      \centering
      \includegraphics[width=0.4\textwidth]{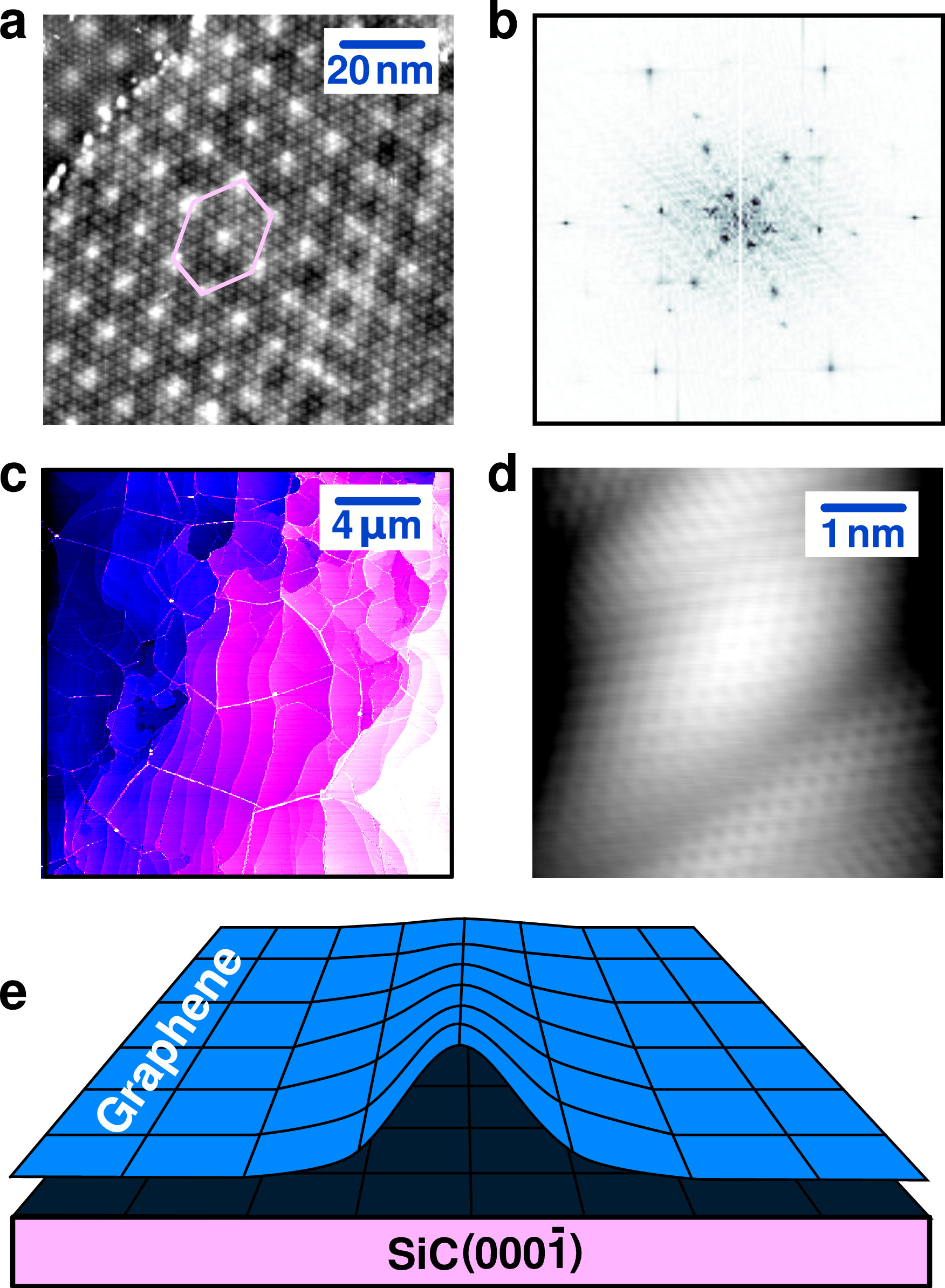}
      \caption %
      {(a) An image of how strain can affect a \moire pattern. (b) displays the FT of the image in (a). As many as 5 sets of \moire spots can be identified, requiring interactions from at least 4 layers. Strain can be caused by low angle grain boundaries seen in (a) or ``pleats''. The AFM image in (c) shows these structures. The pleats are folds in the graphene lattice and tend to run along the SiC step or at angles roughly perpendicular to them. The STM image in (d) shows atomic imaging on top of a pleat. (e) Schematic of a pleat in epitaxial graphene. Mismatches in the thermal contraction after growth can induce buckling in the lattice.}
      \label{fig:MoireStrain}
    \end{figure}

    Given a low enough angle rotation, the relative strain effects are indeed observed. Low angle \moire patterns are even more likely to occur when multiple patterns can be simultaneously observed. Figure \ref{fig:MoireStrain}a shows a fairly large \moire pattern with an average unit cell length of $D = \unit[10.5]{nm} \pm \unit[0.9]{nm}$, corresponding to a rotation angle $\theta = \unit[1.36 \pm 0.11]{^{\circ}}$; however, distortion from relative strain causes the actual size of the \moire pattern to range from $\unit[9.50]{nm}$ to $\unit[11.7]{nm}$, depending on which axis is measured\cite{Error}. This corresponds to a \moire length distortion of $23 \%$ ($D_{max}/D_{min} = 11.7/9.50 = 1.23$); thus we estimate the parameter $\varepsilon \approx 0.37 \%$ for the relative strain between the two lattices producing the pattern.

    There are a few mechanisms by which a graphene sheet can undergo strain. In Fig.~\ref{fig:MoireStrain}a, low angle grain boundaries are visible. The \moire pattern bends with the boundary on the right. In this case, the subsurface rippling of the lattice is likely causing a slight relative strain. Another type of strain possible in graphene grown on \cface is a bending of the graphene sheet caused by unequal thermal contractions between the SiC and graphene sheets after growth \cite{Cambaz08}. In order to relieve stress, the lattice buckles (see Fig.~\ref{fig:MoireStrain}e). The image in Fig.~\ref{fig:MoireStrain}c was taken by atomic force microscopy, and shows these ``pleats'' of the graphene sheet. They are typically $\unit[5]{nm}$ to $\unit[10]{nm}$ in height, and are visible in Fig.~\ref{fig:MoireStrain}c as white lines running parallel and transverse to the SiC steps. Despite bending, the graphene lattice stays in tact over a pleat, and the graphene lattice can extend for several microns before a pleat occurs. The STM image in Fig.~\ref{fig:MoireStrain}d was taken on top of a graphene pleat. The lattice is continuous, but it is clearly contorted as a result of the lattice buckling. The strain reilef looks much like the strain relief in nanotubes\cite{Orlikowski00}. Flat areas near these pleats could show strain effects. Immediately around pleats, slight layer delamination is possible which may not even produce a recognizable \moire pattern to be observed.

\section{Conclusions}
\label{sec:conclusions}

Multilayer epitaxial graphene is known to have carrier mean free times comparable to suspended monolayer graphene\cite{Orlita08,Miller09,Bolotin09}, and measurements to date indicate that its electronic structure is effectively that of independent graphene monolayers\cite{Sadowski07,Sprinkle09a}.  However, the physical structure of MEG is complex.  Using the methods developed here, surface STM imaging can be used to obtain layer alignments and local strain from at least 3 graphene layers in the multilayer stack. These microscopic measurements will be useful for deciphering the formation of rotational domains and for mapping inhomogeneous strain fields, each of which could affect the quantum states of this material---especially under the influence of electric and magnetic fields.

\begin{acknowledgements}
We thank C. Berger, M. Sprinkle, N. Sharma, S. Blankenship, A. Band, and F. Hess for their technical contributions to this work. Funding from NSF (DMR-0804908), the Semiconductor Research Corporation Nanoelectronics Research Initiative (NRI-INDEX), and the W. M. Keck Foundation are gratefully acknowledged. Graphene production facilities were developed under NSF grant ECCS-0521041.
\end{acknowledgements}


\begin{thebibliography}{10}
\bibitem{Error}
All uncertainties reported represent one standard deviation in the measured
  quantity. Error is calculated from the position of the intensity spots of the
  FT.
\end{thebibliography}
\end{document}